\documentclass[11pt]{article}
\usepackage{amsfonts,epsfig,latexsym}

\newcommand{\qed}{\begin{flushright}$\Box$\end{flushright}}

\newfont{\blackb}{msbm10 scaled\magstep1}

\newfont{\calig}{cmsy10 scaled\magstep1}

\def\text#1{\hbox{#1}}

\newtheorem{Theorem}{Theorem}

\newtheorem{Remark}{Remark}

\newtheorem{Lemma}{Lemma}

\def\be{\begin{equation}}
\def\ee{\end{equation}}
\def\ben{\begin{displaymath}}
\def\een{\end{displaymath}}
\def\baa{\begin{eqnarray}}
\def\eaa{\end{eqnarray}}

\def\ba{\begin{array}}
\def\ea{\end{array}}

\def\Ec{{\cal E}}

\def\3{\ss}

\def\l{\lambda}

\def\t{\tau}
\def\th{\vartheta}

\def\phi{\varphi}

\def\C{\mathbb{C}}
\def\Z{\mathbb{Z}}
\def\R{\mathbb{R}}

\def\t0{\Theta_0}

\def\la{\label}

\def\f{\frac}
\def\L{{\cal L}}

\def\tr{{\rm tr}}
\def\0{S}
\def\1{T}

\def\log{\ln}

\def\f{\frac}

\def\8{\infty}

\def\tr{{\rm tr}}

\def\sl2{\mathfrak{sl}(2,\R)}
\def\SL2{SL(2,\R)}

\def\Ot{W}
\def\Om{\Omega}
\def\si{\sigma}

\begin{document}
\begin{center}

{\LARGE Self-dual $SU(2)$ invariant Einstein metrics
and modular dependence  of  theta-functions}
\vskip1.2cm
{\large M.V.Babich\footnote{Permanent address: Steklov Mathematical Institute,
Fontanka, 27, St.Petersburg 191011, Russia}\hskip0.5cm and \hskip0.5cm D.A.Korotkin}
\vskip0.5cm
Max-Planck Institut f\"{u}r Gravitationsphysik\\
Schlaatzweg 1, Potsdam 14473, Germany
\end{center}
\vskip0.5cm
{\bf Abstract}
The goal of this paper is to simplify the  Hitchin's description of
$SU(2)$-invariant self-dual Einstein metrics, making use of the   tau-function
corresponding to a four-pole Schlesinger system.

\vskip1.0cm

The  $SU(2)$ invariant self-dual Einstein metrics were studied in a
number of papers
\cite{Tod94,MaMaWo93,Hitc94}. The local
classification of the metrics
of this type was given in an extensive paper by Hitchin \cite{Hitc94}. 
However, the final form
of  metric coefficients  related to  Painlev\'{e} 6 equation
 was rather complicated in Hitchin's description.

The purpose of this letter it to give simpler expressions for the
same metric,
exploiting the formula for
the tau-function of the algebro-geometric solutions of the  Schlesinger
system found in
the paper \cite{KitKor98}.

Following Tod \cite{Tod91,Tod94}, we start from the following  form of $SU(2)$-invariant self-dual
Einstein metric:
\be
g=F\left\{d\mu^2 +\f{\si_1^2}{W_1^2} +\f{\si_2^2}{W_2^2}+\f{\si_3^2}{W_3^2}\right\}\;,
\la{g}\ee
where 1-forms $\si_j$ satisfy
\be
d\si_1 =\si_2\wedge \si_3\;, \hskip0.5cm
d\si_2 =\si_3\wedge \si_1\;, \hskip0.5cm
d\si_3 =\si_1\wedge \si_2\; ,
\la{ds}\ee
and functions $W_j$ depend only on $\mu$.
Defining the new variables $A_j(\mu)$ by the equations
\be
\f{d W_j}{d\mu} = - W_k W_l + W_j (A_k+ A_l)\;,
\la{motion}
\ee
where  $(j,k,l)$ are arbitrary permutations of indexes $(1,2,3)$, we can write the
condition of self-duality of related manifold   
in the form of the classical Halphen system:
\be
\f{d A_j}{d\mu} = - A_k A_l + A_j (A_k+ A_l)\;.
\la{Halphen}\ee
Once the system (\ref{Halphen}) is solved, 
the solution may be substituted into the system (\ref{motion}), which defines then 
metric coefficients $W_j$.

The system (\ref{motion}), (\ref{Halphen}) is invariant 
with respect to $SL(2,\R)$ M\"{o}bius transformations:
\be
\mu\to \tilde{\mu} \f{a\mu + b}{c\mu + d}\;,\hskip1.0cm ad-bc=1 \;,\hskip0.6cm
\la{conmu}\ee
\be
W_j\to  \tilde{W}_j = (c\mu+d)^2 W_j\;,\hskip2.0cm
\la{conW}\ee
\be
A_j\to  \tilde{A}_j=  (c\mu+d)^2 A_j +c (c\mu+d)^2\;.
\la{conA}\ee

If one chooses the trivial solution $A_j=0$ of the system (\ref{Halphen}), the  corresponding  system  (\ref{motion}) 
reduces to equations of the Euler top; this case was considered in \cite{BGPP78,EguHan78}.
If two of the  functions $A_j$ vanish, the remaining one must be  constant; this case was treated in the 
paper  \cite{PedPoo90}. Another case, when for all $j$ one chooses $W_j=A_j$, was discussed in \cite{AtiHit88}.

The system (\ref{motion}), corresponding to a  general solution of the
system  (\ref{Halphen}), was related in the papers \cite{Tod94,Hitc94} to the four-point 
Schlesinger system. Moreover, it turned out, that the conformal factor $F$ 
may be chosen to make metric (\ref{g}) satisfy the Einstein equation exactly in the case when the
system (\ref{motion}) can be solved in elliptic functions. However, if one would like to
extract explicit formulas for coefficients $W_j$ from the paper \cite{Hitc94}, the result 
turns out to be rather complicated. 

The purpose of this paper is to derive  simple 
formulas for functions $W_j$ and $F$, corresponding to the general solution 
of  (\ref{Halphen}), using the formula for the $\tau$-function of the Schlesinger system 
found in the paper \cite{KitKor98}.

Let us define the standard theta-function with characteristics by the series
\ben
\th[p,q](z,\sigma)= \sum_{m\in\Z}\exp\{\pi i (m+p)^2\sigma+2\pi i (m+p)(z+q)\}\;,\hskip 1.5cm
z,\sigma,p,q\in\C\;;\;\; \Im\sigma>0
\een

It is well-known (see, for example, \cite{HarMcK98,ChAbCl90,Takh93}), that the functions
\be
A_1= 2\f{d}{d\mu}\log\th_2\;,\hskip0.5cm 
A_2= 2\f{d}{d\mu}\log\th_3\;,\hskip0.5cm A_3= 2\f{d}{d\mu}\log\th_4\;,
\la{Wtheta}\ee 
where
\be
\th_2 \equiv \th\Big[\f{1}{2},0\Big](0,i\mu)\;, \hskip0.5cm
\th_3 \equiv \th[0,0](0,i\mu)\;, \hskip0.5cm
\th_4 \equiv \th\Big[0,\f{1}{2}\Big](0,i\mu)\;
\la{Jacobi}\ee
are standard theta-constants, solve the Halphen system (\ref{Halphen}). 
 The general solution of (\ref{Halphen}) may be obtained applying M\"{o}bius transformations 
(\ref{conmu}), (\ref{conA}) to solution (\ref{Wtheta}). Therefore, it is sufficient to 
solve the system  (\ref{motion}), where functions  $A_j$ are given by  (\ref{Wtheta}). Then the solution of 
 the system  (\ref{motion}), corresponding to  the general solution of the Halphen system 
(\ref{Halphen}) may be  obtained by applying a  M\"{o}bius transformation (\ref{conW}). 
Thus  in the sequel we shall work  with the following non-autonomous system of equations 
\be
\f{dW_1}{d\mu}= - W_2 W_3+ 2 W_1\f{d}{d\mu}\log(\th_3\th_4)\; ,
\la{eqOt1}\ee
\be
\f{dW_2}{d\mu}= - W_3 W_1+ 2 W_2\f{d}{d\mu}\log(\th_2\th_4)\; ,
\la{eqOt2}\ee
\be
\f{dW_3}{d\mu}= - W_1 W_2+ 2 W_3\f{d}{d\mu}\log(\th_2\th_3)\; .
\la{eqOt3}\ee

It was found by Tod \cite{Tod94} that for a special class of solutions of this system
the conformal factor $F$ in (\ref{g}) can be chosen in such a way that the
metric $g$ satisfies   Einstein's equations (with cosmological constant).
 In terms of $\{W_j\}$ this class of solutions is characterised by the following condition:
\be
\th_2^4\Ot_1^2-\th_3^4\Ot_2^2+\th_4^4\Ot_3^2=\f{\pi^2}{4}\th_2^4\th_3^4\th_4^4\; .
\la{intt}\ee

The rest of the letter is devoted to the proof of the following theorem:

\begin{Theorem}
The general two-parametric family of solutions of the system 
(\ref{eqOt1}),  (\ref{eqOt2}),  (\ref{eqOt3}), satisfying condition
(\ref{intt}), is given by the following formulas:
$$
W_1=-\f{i}{2}\th_3\th_4\f{\f{d}{d q}\th[p,q+\f{1}{2}]}{e^{\pi i p}\th[p,q]}\;,\hskip1.0cm
W_2=\f{i}{2}\th_2\th_4\f{\f{d}{d q}\th[p+\f{1}{2},q+\f{1}{2}]}{e^{\pi i p}\th[p,q]}\;,$$
\be
W_3=-\f{1}{2}\th_2\th_3\f{\f{d}{d q}\th[p+\f{1}{2},q]}{\th[p,q]}\;,
\la{WWW}\ee
where $\th[p,q]$ denotes the  theta-function with characteristics
of vanishing first argument $\th[p,q](0,i\mu)$; $p,q\in\C$. 

The corresponding metric (\ref{g}) is real and satisfies  Einstein's equations with  a negative cosmological constant $\Lambda$
if 
\be
p\in\R\;, \hskip0.8cm \Re q=\f{1}{2}
\la{rc2}\ee
and the conformal factor $F$ is given by the following formula:
\be
F= \f{2}{\pi\Lambda} \f{W_1 W_2 W_3}{\Big(\f{d}{dq}\log\th[p,q]\Big)^2}\;.
\la{Fgen}\ee
The metric (\ref{g}) is real and satisfies  Einstein's equations with a positive cosmological constant $\Lambda$
if 
\be
q\in\R\;, \hskip0.8cm \Re p=\f{1}{2}\;,
\la{rc4}
\ee
and the conformal factor is given by the same formula (\ref{Fgen}).

There exists also an additional one-parametric family of solutions of the system  (\ref{eqOt1}), (\ref{eqOt2}),  (\ref{eqOt3}), (\ref{intt}):
\be
W_1=\f{1}{\mu+ q_0}+ 2\f{d}{d\mu}\log\th_2 \;,\hskip0.7cm
W_2=\f{1}{\mu+ q_0}+ 2\f{d}{d\mu}\log\th_3 \;,\hskip0.7cm
W_3=\f{1}{\mu+ q_0}+ 2\f{d}{d\mu}\log\th_4 \;
\la{WWWdeg}\ee
 ($q_0\in\R$), which defines  manifolds with the vanishing cosmological constant $\Lambda=0$
if the  conformal factor $F$ is defined by the formula
\be
F= C (\mu+q_0)^2 W_1 W_2 W_3\;,
\la{Fdeg}\ee
where $C>0$ is an arbitrary constant.
\end{Theorem}

\vskip0.6cm
Let us now introduce auxiliary 
dependent variables $\Om_j$ via the following relations:
\be
\Om_1 =  -\f{W_2}{\pi \th_2^2 \th_4^2} \; ,\hskip0.5cm
\Om_2 =  -\f{W_3}{\pi \th_2^2 \th_3^2} \; ,\hskip0.5cm
\Om_3 =  -\f{W_1}{\pi \th_3^2 \th_4^2} \;.
\la{Om}\ee
(we performed the cyclic permutation of indexes $(1,2,3)$ to bring our notations in agreement with
the papers \cite{Tod94,Hitc94}).

In terms of variables  $\Om_j$ the system (\ref{eqOt1}) - (\ref{eqOt3}) looks as follows:
\be
\f{d\Om_1}{d\mu}= \pi \th_{3}^4 \Om_2 \Om_3\; ,\hskip0.5cm
\f{d\Om_2}{d\mu}= \pi \th_{4}^4 \Om_3 \Om_1\; ,\hskip0.5cm
\f{d\Om_3}{d\mu}= \pi \th_{2}^4 \Om_1 \Om_2\;.
\la{auxsys}\ee
The condition (\ref{intt}) now takes the simple form
\be
-\Om_1^2+\Om_2^2+\Om_3^2=\f{1}{4}\;.
\la{int}\ee

Let us also introduce some other auxiliary objects.
Consider an elliptic curve $\L$ defined by the equation
\be
\nu^2= \l(\l-1)(\l-x)\;.
\la{L}\ee
Assume that  $x$ varies between $0$ and $1$, and choose the
basic $a$-cycle to encircle branch cut $[0,x]$, and
$b$-cycle to encircle branch points $x$ and $1$.
We shall introduce  the full elliptic integral $K$ along $a$-cycle:
\be
K=\f{1}{2}\int_0^x\f{d\l}{\sqrt{\l(\l-1)(\l-x)}}\;.
\la{w}\ee
Then we can define the $b$-period  $i\mu$ ($\mu\geq 0$) of the curve
$\L$ as follows:
\be
i\mu=\f{1}{2K}\int_x^1\f{d\l}{\sqrt{\l(\l-1)(\l-x)}}\;.
\la{ka}\ee
Thre relation (\ref{ka}) defines the one-to-one correspondence between 
variables $x$ and $\mu$; the  dependence
$\mu(x)$ is described by the following equation:
\be
\f{d\mu}{d x}=\f{\pi }{4K^2 x(x-1)}\;.
\la{sx}\ee
We can express theta-constants in terms of $x$ and $w$ as follows:
\be
\th_2^4 =  \f{4}{\pi^2} K^2 x\;, \hskip0.8cm
\th_3^4 =  \f{4}{\pi^2} K^2 \;, \hskip0.8cm
\th_4^4 =  \f{4}{\pi^2} K^2 (1-x)\;.
\la{thomae}\ee
Choosing $x$ as an independent variable, one can rewrite the system  (\ref{auxsys})  in a more familiar form:
\be
\f{d\Om_1}{d x}= -\f{\Om_2\Om_3}{x(1-x)}\;, \hskip0.5cm
\f{d\Om_2}{d x}= -\f{\Om_3\Om_1}{x}\;, \hskip0.5cm
\f{d\Om_3}{d x}= -\f{\Om_1\Om_2}{1-x}\;. \hskip0.5cm 
\la{eqOm}\ee

The metric (\ref{g}) in terms of the new variables takes the following form:
\be
g=\tilde{F}\left\{\f{dx^2}{x(1-x)} +\f{\si_1^2}{\Om_1^2} +\f{(1-x)\si_2^2}{\Om_2^2}+
\f{x\si_3^2}{\Om_3^2}\right\}\;,
\la{g1}\ee 
where the ``new'' conformal factor $\tilde{F}$ is related to the ``old'' conformal factor $F$ as follows:
\be
\tilde{F}=\f{1}{\th_2^4\th_4^4} F\;.
\la{FFt}\ee

If variables $\Om_j$ solve the system (\ref{eqOm}),
and satisfy condition (\ref{int}), then the 
metric (\ref{g1}) satisfies  Einstein's equations with cosmological constant  $\Lambda$
 if the factor $\tilde{F}$ is given by the
following expression \cite{Tod94}:
\be
\tilde{F}=-\f{1}{4 \Lambda}\f{8x\Om_1^2\Om_2^2\Om_3^2 + 2\Om_1\Om_2\Om_3
(x(\Om_1^2+\Om_2^2)-(1-4\Om_3^2)
(\Om_2^2-(1-x)\Om_1^2))}{(x\Om_1\Om_2+2\Om_3(\Om_2^2-(1-x)\Om_1^2))^2}\;.
\la{F}\ee

\vskip0.5cm

The system (\ref{eqOm}) arises in the context of isomonodromic deformations
of the ordinary matrix
differential equation {}
\be
\f{d\Psi}{d\l} = \Big(\f{A^0}{\l}+\f{A^1}{\l-1}+\f{A^x}{\l-x}\Big)\Psi\;,
\la{ls}\ee
where $A^0,\,A^1,\,A^x\in sl(2,\C)$; $\Psi(\l)\in SL(2,\C)$.

The condition of $x$-independence of monodromies, $M^0$, $M^1$ and
$M^x$, of the function $\Psi$ around
points $0$, $1$ and $x$, together with the assumption of $x$-independence of
the main term of the asymptotical expansion of
$\Psi$ at $\l=\infty$ (which may be always achieved by a gauge
transformation $\Psi\to C(x)\Psi$), imply the following
dependence of $\Psi$ on the variable $x$:
\be
\f{d\Psi}{d x}= - \f{A^x}{\l-x}\Psi\;.
\la{lsx}\ee
The compatibility condition of equations (\ref{ls}) and (\ref{lsx}) is
equivalent to the following  Schlesinger system:
\be
\f{dA^0}{dx}=\f{[A^x,A^0]}{x}\;, \hskip0.5cm
\f{dA^1}{dx}=\f{[A^x,A^1]}{x-1}\;, \hskip0.5cm
\f{dA^x}{dx}=-\f{[A^x,A^0]}{x}-\f{[A^x,A^1]}{x-1}\;.
\la{Sch}\ee
The problem of finding function $\Psi$ and residues $A_j$ corresponding
to a given set of monodromy matrices
in called the inverse monodromy problem (or Riemann-Hilbert problem).
Notice that the transformation
of function
\be
\Psi\to \Psi S\;,
\ee
 where $S$ is constant invertible matrix, leaves corresponding solution
$\{A_j\}$ of the Schlesinger system invariant,
and transforms monodromy matrices $M_j$ as follows:
\be
M_j\to S^{-1} M_j S\;.
\la{simM}\ee

If we fix values of the integrals of motion of the system (\ref{Sch}) in the following way:
\be
 \tr (A^0)^2=\tr (A^1)^2=\tr (A^x)^2=\f{1}{8}\;,
\la{trA}\ee
 the formulas
\be
\Om_1^2=-(\f{1}{8} +\tr A^0 A^1)\;, \hskip0.5cm
\Om_2^2=\f{1}{8} +\tr A^1 A^x\;, \hskip0.5cm
\Om_3^2=\f{1}{8} +\tr A^0 A^x
\la{defOm}\ee
give a solution of  the system (\ref{eqOm}), (\ref{int}) (see \cite{Hitc94,Fokas}).

As a corollary of conditions  (\ref{trA}), eigenvalues of all
monodromies $M^0$, $M^1$ and $M^x$ equal $\pm i$.
Such sets of monodromy matrices allow simple classification
\cite{Hitc94}:
\begin{Theorem}
Let eigenvalues of all monodromy matrices $M^0$, $M^1$ and $M^x$ equal
$\pm i$. Then up to a  simultaneous constant
similarity transformation they are given by
$$                                                 {}
M^0=\left(\ba{cc}0 & -ie^{-2\pi i q} \\
                 -ie^{2\pi i q} & 0      \ea\right)\;; \hskip0.8cm
M^1=\left(\ba{cc}0 & ie^{-2\pi i (p+q)} \\
                 i e^{2\pi i (p+q)} & 0      \ea\right)\;;
$$
\be
M^x=\left(\ba{cc}0 & -ie^{-2\pi i p} \\
                 -i e^{2\pi ip} & 0      \ea\right)\; \hskip0.8cm
p,q\in \C\;,
\la{genM}\ee
or
\be
M^0=\left(\ba{cc}  -i & q_0 \\
                    0 & i     \ea\right)\;; \hskip0.5cm
M^1=\left(\ba{cc} -i  & -i+ q_0   \\
                   0  & i      \ea\right)\;; \hskip0.5cm
M^x=\left(\ba{cc} -i  & -i \\
                   0  & i     \ea\right)\;, \hskip0.5cm  q_0\in \C
\,.                                       {}
\la{degM}\ee
\end{Theorem}

It is easy to see, however, that the solution of the inverse monodromy
problem for
one-parametric set of monodromy matrices
(\ref{degM}) may be obtained in certain limit of solution of the
inverse monodromy problem for
the two-parametric set of monodromy matrices (\ref{genM}). Namely, we
can formulate the following

\begin{Lemma}
Let functions $A_j(x,p,q)$ give a solution of Schlesinger system
(\ref{Sch}) corresponding to monodromy
matrices (\ref{genM}). Then the solution of the Schlesinger system
corresponding to monodromy
matrices (\ref{degM}) is given by
\be
A_j(x,q_0)= \lim_{\epsilon\to 0}
A_j(x,\f{1}{2}+\epsilon,\f{1}{2}+i q_0\epsilon)\;.
\ee
\la{Lemmadeg}
\end{Lemma}
{\it Proof.} Let us put in monodromy matrices  (\ref{genM})
$p=\f{1}{2}+\epsilon$, $q=\f{1}{2}+i q_0\epsilon$
and apply to these matrices the simultaneous similarity transformation
(\ref{simM}) with matrix $S$ given by
\be
S=\left(\ba{cc}   4\pi i    &   \epsilon^{-1}  \\
                    0       &   \epsilon^{-1}    \ea\right)\;.
\la{K}\ee
Then in the limit $ \epsilon\to 0$ we come to the set of monodromy
matrices  (\ref{degM}). Taking into account that the
solution of the Schlesinger system is invariant with respect to
simultaneous  transformation (\ref{simM})
of all monodromy matrices, we see that the same limit on the level of
matrices $A_j$ gives solution
corresponding to monodromies  (\ref{degM}).
\qed
This lemma shows that it is sufficient to restrict ourselves to
solution of the inverse monodromy problem
with the two-parametric set of monodromy matrices  (\ref{genM}); then the
solution corresponding to monodromy matrices
(\ref{degM}) may be found by the  simple limiting procedure.

In \cite{Hitc94} the system (\ref{eqOm}), (\ref{int})  was solved in terms
of elliptic functions
(independently in the cases (\ref{genM}) and  (\ref{degM}))
by exploiting the link between $\{\Om_j\}$ and the solution $y(x)$ of the Painlev\'{e} 6
equation with coefficients
$(\f{1}{8},\,-\f{1}{8},\,\f{1}{8},\,\f{3}{8})$, which is known to be
equivalent to the system (\ref{Sch}).
In principle, this solution could also be extracted from the earlier paper of
Okamoto \cite{Okam87}.
If one  then  directly expresses $\{\Om_j\}$ in terms of elliptic
functions according to the Hitchin's
or Okamoto's schemes, the resulting expressions turn out to be very
cumbersome.

Solutions of the Painlev\'e 6 equation
corresponding to monodromy matrices (\ref{genM}) and (\ref{degM}) are
of course also related by the limiting procedure described in the
previous lemma.

Now we are going to give an alternative simpler
formulas for $\Om_j$,
making use of the Jimbo-Miwa tau-function, $\tau(x)$, of the Schlesinger system
(\ref{Sch} which is  defined by the following
equation \cite{MiwJim81}:
\be
\f{d}{dx}\log\tau (x)= \f{\tr A^0 A^x}{x} + \f{\tr A^1 A^x}{x-1}\;.
\la{taudef}\ee

The solution of the system
(\ref{eqOm}) 
can be expressed in terms of the
$\tau$-function (\ref{taudef}) as follows:

\begin{Theorem}
Let the tau-function  $\tau(x)$ be defined by (\ref{tau}). Then functions
\be
\Om_1^2=\f{d}{dx}\{x(x-1)\f{d}{dx}\log\tau(x)\}\;,
\la{Om1}\ee
\be
\Om_2^2=(1-x)\f{d}{dx}\{x(x-1)\f{d}{dx}\log\tau(x)\}+x(x-1)\f{d}{dx}\log\tau(x)+\f{1}{8}\;,
{}
\la{Om2}\ee
\be
\Om_3^2=x\f{d}{dx}\{x(x-1)\f{d}{dx}\log\tau(x)\}-x(x-1)\f{d}{dx}\log\tau(x)+\f{1}{8}
\la{Om3}\ee
solve   (\ref{eqOm}), (\ref{int}).
\end{Theorem}

{\it Proof.} According to the definition (\ref{taudef}) of the
$\tau$-function, and constraint (\ref{int}), we have
\be
x(x-1)\f{d}{dx}\log\tau = (x-1)(\Om_3^2-\f{1}{8}) + x (\Om_2^2-\f{1}{8})
= (x-1)\Om_1^2+\Om_2^2-\f{1}{8}\;.
\la{cal1}\ee
Using equations (\ref{eqOm}), we can calculate the $x$-derivative of the
right hand side of this equation:
\be
\f{d}{dx}\Big((x-1)\Om_1^2+\Om_2^2-\f{1}{8}\Big)=
\Om_1^2+2(x-1)\Om_1\Big(-\f{\Om_2\Om_3}{x(1-x)}\Big)+2\Om_1\Big(-\f{\Om_1\Om_3}{x}\Big)=\Om_1^2\;,\ee
which implies (\ref{Om1}). Expressions  (\ref{Om2}) and  (\ref{Om3})
follow now from relation
(\ref{cal1}) if we take into account  the fixed value of the integral of
motion (\ref{int}).

\qed

Now we are in position to formulate the following theorem:
\begin{Theorem}
The $\tau$-function of the Schlesinger system (\ref{Sch}), corresponding
to monodromy matrices (\ref{genM}), is given by
the following expression:
\be
\tau(x)=\f{\th[p,q]}{\sqrt{\th_2\th_4}}\;.
\la{tau}
\ee
The $\tau$-function, corresponding to monodromy matrices  (\ref{degM}),
is given by the formula
\be
\tau(x)=(\mu(x)+q_0)\th_2\sqrt{\th_2\th_4}\;.
\la{taudeg}\ee
\la{theotau}
\end{Theorem}

{\it Proof.} In \cite{KitKor98} it was proved the following formula for
the $\tau$-function,
corresponding to monodromy matrices  (\ref{degM}):
\be
\tau(x)=\f{\th[p,q]}{K^{1/2}[x(x-1)]^{1/8}}\;,
\ee
which turns into (\ref{tau}) if we take into account the expressions for
the theta-constants (\ref{thomae})
in terms of $K$ and $x$.

Now, taking into account Lemma \ref{Lemmadeg},  we can get
the $\tau$-function, corresponding to monodromy matrices  (\ref{degM})
 by choosing in (\ref{tau}) $p= 1/2+\epsilon$,
$q= 1/2+q_0\epsilon$ and taking the limit  $\epsilon\to 0$. The result, up to a
non-essential $x$-independent factor, turns out to look as follows:
$$
\tau(x)=\f{(\mu+q_0)\th_1'(0)}{K^{1/2}[x(x-1)]^{1/8}}\;,
$$
where prime denotes the derivative with respect to the first argument of theta-function.
To complete the proof of expresion (\ref{taudeg}) it remains to use the
Jacobi formula
$$
\th_1'(0)= \pi \th_2\th_3\th_4\;.
$$
where we adopt the sign convention $\th_1(z)\equiv -\th[\f{1}{2},\f{1}{2}](z)$.
\qed

 Substitution of $\tau$-function  (\ref{tau}) into expressions
(\ref{Om}), leads to the following theorem:
\begin{Theorem}
 Solution of the system (\ref{eqOt1}), (\ref{eqOt2}),
(\ref{eqOt3}),
corresponding to monodromy matrices  (\ref{genM}), is given by the following formulas:
\be
W_1^2=-\f{d^2}{d\mu^2}\log\f{\th[p,q]}{\sqrt{\th_3\th_4}} +
4\Big\{\f{d}{d\mu}\log\th_2\Big\}\Big\{\f{d}{d\mu}\log\f{\th[p,q]}{\sqrt{\th_3\th_4}}\Big\}\; ,
\la{solOt1}\ee
\be
W_2^2=-\f{d^2}{d\mu^2}\log\f{\th[p,q]}{\sqrt{\th_2\th_4}} +
4\Big\{\f{d}{d\mu}\log\th_3\Big\}\Big\{\f{d}{d\mu}\log\f{\th[p,q]}{\sqrt{\th_2\th_4}}\Big\}\; ,
\la{solOt2}\ee
\be
W_3^2=-\f{d^2}{d\mu^2}\log\f{\th[p,q]}{\sqrt{\th_2\th_3}} +
4\Big\{\f{d}{d\mu}\log\th_4\Big\}\Big\{\f{d}{d\mu}\log\f{\th[p,q]}{\sqrt{\th_2\th_3}}\Big\}\; .
\la{solOt3}\ee
\end{Theorem}

Remarkably, it turns out to be possible to calculate the square roots of these expressions to get the
simple formulas for $W_j$  themselves. The result is given by the following lemma:

\begin{Lemma}\la{lemW}
 The square roots of expressions (\ref{solOt1})- (\ref{solOt3}), satisfying the  system (\ref{eqOt1}) - (\ref{eqOt3}),
 are given by the following formulas:
\be
W_1=-\f{i}{2}\th_3\th_4\f{\f{d}{d q}\th[p,q+\f{1}{2}]}{e^{\pi i p}\th[p,q]}\;,
\la{ident1}\ee
\be
W_2=\f{i}{2}\th_2\th_4\f{\f{d}{d q}\th[p+\f{1}{2},q+\f{1}{2}]}{e^{\pi i p}\th[p,q]}\;,
\la{ident2}\ee
\be
W_3=
-\f{1}{2}\th_2\th_3\f{\f{d}{d q}\th[p+\f{1}{2},q]}{\th[p,q]}\;.
\la{ident3}\ee

\end{Lemma}
The proof of the lemma may be obtained via lenghy
elementary calculations expressing both sides of these identities via
elliptic functions $sn(u)$, $cn(u)$, $dn(u)$ (where $u=2K(p\sigma+q+\f{1}{2})$) 
and standard complete elliptic integrals $K(x)$ and $E(x)$ of the first and second kind, respectively. 
It is also convenient to introduce the notations 
$$\Ec=E/K \hskip0.5cm {\rm and} \hskip0.5cm Z= \f{1}{2K}\f{d}{dq}\log\th(p\sigma+ q)\;. $$
The $x$-dependence of $K$ and $\Ec$ is described by the following equations:
\ben
\f{d}{dx}\log K^2=\f{\Ec}{x(1-x)}-\f{1}{x}\;,\hskip1.0cm 
2\f{d}{dx}\log\Ec=-\f{\Ec^2}{x(1-x)}+\f{2\Ec}{x}-\f{1}{x}\;.
\een
The $\mu$ - derivatives of the theta-constants may be expressed as follows:
\ben
\f{d}{d\mu}\log\th_2 = -\f{K^2}{\pi}\Ec\;,\hskip0.7cm 
\f{d}{d\mu}\log\th_3= -\f{K^2}{\pi}(\Ec-1+x)\;,\hskip0.7cm
\f{d}{d\mu}\log\th_4= -\f{K^2}{\pi}(\Ec-1)\;
\een
and 
\ben
\f{d^2}{d\mu^2}\log\th_2=\f{2K^4}{\pi^2}(\Ec^2-1+x)\;,\hskip1.0cm 
\f{d^2}{d\mu^2}\log\th_3=\f{2K^4}{\pi^2}(\Ec^2-2\Ec(1-x)-x+1)\;,\hskip1.0cm 
\een
\ben
\f{d^2}{d\mu^2}\log\th_4=\f{2K^4}{\pi^2}(\Ec^2-2\Ec+1-x)\;.
\een
Finally, taking into account the expression
\ben
dn^2(u)=\f{d^2}{du^2}\log\th_4\Big(\f{u}{2K}\Big)+\Ec\;,
\een
and standard relations between elliptic functions:
\ben
\f{d\, sn(u)}{du}= cn(u) dn(u)\;,\hskip1.0cm  \f{d\,cn(u)}{du}= -sn(u) dn(u)\;,\hskip1.0cm
 \f{d\,dn(u)}{du}=-x\, sn(u) cn(u)\;,
\een
\ben
sn^2(u)+ cn^2(u)=1 \;,\hskip1.0cm
x sn^2(u) + dn^2(u) =1\;,
\een
we obtain for $q$-derivatives of variable $Z$:
\ben
\f{d}{d q} \{2 K Z\}= (2K)^2 (dn^2(u)-\Ec)\;,\hskip1.0cm 
\f{d^2}{d q^2} \{2 K Z\}=-2x (2K)^3 sn(u) cn(u) dn(u)\;,
\een
\ben
\f{d^2}{d q^3} \{2 K Z\}= 2x (2K)^4 \{2x sn^2(u)cn^2(u)+ sn^2(u) dn^2(u)+cn^2(u)\}\;.
\een  
Now substitution of all these auxiliary formulas 
into the expressions (\ref{solOt1}) - (\ref{solOt3}) for $W_j^2$  gives the following result:
\ben
W_1^2= -\f{(2K)^4}{4\pi^2}[x sn(u)cn(u)- Z dn(u)]^2\;,\hskip1.0cm
W_2^2= -\f{x(2K)^4}{4\pi^2}[sn(u) dn(u) -Z cn(u)]^2\;,
\een
\ben 
W_3^2= \f{x(2K)^4}{4\pi^2}[cn(u)dn(u)+ Z sn(u)]^2\;.
\een
These expressions
 (up to the choice of the sign) coincide with  (\ref{ident1}) - (\ref{ident3}) if we take into 
account the expressions
for $sn$, $cn$ and $dn$ in terms of theta-functions:
\ben
sn(u) = \f{\th_3}{\th_2}\f{\th_1(u/2K)}{\th_4(u/2K)}\;,\hskip1.0cm
cn(u) = \f{\th_4}{\th_2}\f{\th_2(u/2K)}{\th_4(u/2K)}\;,\hskip1.0cm
dn(u) = \f{\th_4}{\th_3}\f{\th_3(u/2K)}{\th_4(u/2K)}\;.
\een
\qed

The signs in the formulas (\ref{ident1}) - (\ref{ident3}) are chosen to make $W_j$ 
coincide with solution $W_j=A_j$ in appropriate limit, described by the following lemma:

\begin{Lemma}\la{LemdegW}
Let $p=\f{1}{2}+\epsilon\;,\; q=\f{1}{2}+i q_0\epsilon$. Then in the limit $\epsilon\to 0$
functions $W_j$ given by  (\ref{ident1}) - (\ref{ident3}) turn into
\be
W_1=\f{1}{\mu+ q_0}+ 2\f{d}{d\mu}\log\th_2 \;,\hskip0.7cm
W_2=\f{1}{\mu+ q_0}+ 2\f{d}{d\mu}\log\th_3 \;,\hskip0.7cm
W_3=\f{1}{\mu+ q_0}+ 2\f{d}{d\mu}\log\th_4 \;.
\la{Wdl}\ee
In the limit $q_0\to \infty$ these expressions turn into  solution $W_j=A_j=2\f{d}{d\mu}\log\th_{j+1}$. 
\end{Lemma}
{\it Proof.}  As $\epsilon\to 0$, both the numerator and denominator of each of the expressions 
 (\ref{ident1}) - (\ref{ident3}) vanish. To apply the l'Hopital rule, for example, to $W_1$,
we find:
\ben
\f{d}{d\epsilon} \left\{e^{-\pi i(p+\f{1}{2})}\f{d}{dq}\th[p,q+\f{1}{2}]\right\}(\epsilon=0)=
-i(\mu+q_0)\th_2''(0)+2\pi i \th_2 \;,
\een
\ben
\f{d}{d\epsilon}\th[p,q] (\epsilon=0)= -i(\mu+q_0)\th_1'(0)
\een
which leads to the expression for $W_1$ stated in the lemma after applying the heat equation and the Jacobi formula
for $\th_1'(0)$. Analogously we get the limits of $W_2$ and $W_3$. 
In particular, this confirms that the
signs in (\ref{ident1}) - (\ref{ident3}) are chosen in such a way that functions $W_j$ satisfy the system
(\ref{eqOt1})-(\ref{eqOt3}).
\qed
\begin{Remark}\rm
From the fact that the functions (\ref{Wdl}) satisfy the system (\ref{eqOt1})-(\ref{eqOt3})
for $q_0=\infty$, when
\ben
W_j=A_j=2\f{d}{d\mu}\log \th_{j+1}\;,
\een
 a trivial direct calculation is needed to check that they satisfy the system
for arbitrary $q_0$.
\end{Remark}

The next lemma gives the formula for the conformal facor $F$.
\begin{Lemma}
Conformal factor $F$, defined by the formulas (\ref{F}), (\ref{FFt}) and (\ref{Om}), with functions 
$W_j$ given by the formulas (\ref{ident1}) - (\ref{ident3}), may be represented as follows:
\be
F = \f{2}{\pi\Lambda} \f{W_1 W_2 W_3}{\Big(\f{d}{dq}\log\th[p,q]\Big)^2}
\la{conlem}
\ee
The conformal factor corresponding to solution (\ref{WWWdeg}) may be obtained from this
expression in the limit $\epsilon\to 0$ assuming $p=\f{1}{2}+\epsilon$, $p=\f{1}{2}+i q_0\epsilon$,
$\f{2}{\pi\Lambda}= C\epsilon^{-2}$ with an arbitrary $C>0$. This leads to
\ben
F= C(\mu + q_0)^2 W_1W_2 W_3
\een
In turn, in the limit $q_0\to \infty$, choosing $C=\f{\tilde{C}}{q_0^2}$, we get the conformal factor for
solution $W_j=A_j$:
\ben
F= \tilde{C}  W_1W_2 W_3
\een
\end{Lemma}
The proof of expression 
(\ref{conlem}) again has purely calculational character. Notice just that, as an intermediate step, it is
convenient to rewrite expression (\ref{F}) as follows:
\ben
\tilde{F}= -\f{\Om_1\Om_2\Om_3}{2\Lambda}\f{x(2\Om_1\Om_3+\Om_2)^2- (2\Om_3^2-\f{1}{2})^2}
{(x\Om_1\Om_2+2\Om_3(\Om_2^2-(1-x)\Om_1^2))^2}
\een
The rest of the calculations is again fulfilled in terms of the functions 
$sn$, $cn$, $dn$, $K$, $\Ec$ and $Z$.
\qed

The reality  of  functions $W_j^2$  
(\ref{solOt1}) -  (\ref{solOt3}) is  provided by the following condition:
\be
\f{d}{d\mu}\log\th_{pq}\equiv -\pi p^2 + \f{d}{d\mu}\log\th(i\mu p +q)   \in\R\;,
\la{rc}\ee
which is fulfilled in the following four cases: 
\be
p\in \R\;,\hskip1.0cm  \Re q=\f{1}{2}\;,
\la{rc10} 
\ee
\be
q\in \R\;,\hskip1.0cm  \Re p=\f{1}{2}\;,
\la{rc20}
\ee
\be
p\in \R\;,\hskip1.0cm  \Re q= 0\;,
\la{rc30}
\ee
\be
q\in \R\;,\hskip1.0cm  \Re p= 0\;.
\la{rc40}
\ee
Using  Lemma \ref{lemW}, one can veryfy that in the cases (\ref{rc10}) and 
(\ref{rc20}) functions $W_j$ themselves are also real. Whereas in each of the cases  (\ref{rc30}) and  
(\ref{rc40}) two of the functions $W_j$ turn out to be imaginary.

According to Hitchin's classification, parameters $p$ and $q$,  satisfying reality conditions (\ref{rc10}),
correspond to the self-dual Einstein manifolds with negative scalar curvature. Parameters,  
satisfying reality conditions (\ref{rc20}), correspond to  the manifolds with positive scalar
curvature.

The family of solutions (\ref{Wdl}),
which are real if $q_0$ is real, corresponds to manifolds with vanishing scalar curvature.

This ends the proof of the theorem 1.
\qed

Let us now briefly translate the properties of the metric  (\ref{g})
(for the case of the general solution (\ref{WWW})), investigated in \cite{Hitc94}, to our language.
The metric  (\ref{g})
can be rewritten as follows 
after substitution of expression for conformal factor (\ref{Fgen}): 
\be
ds^2= \f{2}{\pi\Lambda}\left\{\f{\th[p,q]}{\f{d}{d q}\th[p,q]}\right\}^2
\left\{W_1 W_2 W_3\, d\mu^2+ \f{W_2 W_3}{W_1}\si_1^2 + \f{W_1 W_3}{W_2}\si_2^2
+ \f{W_1 W_2}{W_3}\si_3^2\right\}\;.
\la{ggg}\ee
From the expressions (\ref{WWW}) we see 
that  the metric may become singular at whose values of $\mu$, where at least one of the 
following functions:
\be
\th[p,q]\;,
\la{sin1}\ee
\be
\f{d}{d q}\th[p,q]\;,
\la{sin2}\ee
\be
\f{d}{d q}\th[p,q+\f{1}{2}]\;,
\la{sin3}\ee
\be
\f{d}{d q}\th[p+\f{1}{2},q]\;,
\la{sin4}\ee
\be
\f{d}{d q}\th[p+\f{1}{2},q+\f{1}{2}]\;,
\la{sin5}\ee
vanish.
\begin{Remark}\rm
A simple comparison with Hitchin's treatment in terms of the solution of Painlev\'{e} 6 equation
$y(x)$  shows that the zeros of $\th[p,q]$ correspond, up to the map $x\to \mu(x)$,
 to the poles of $y(x)$ with
positive residues (these values Hitchin denoted by   $x_j$). The  
zeros of $\f{d}{d q}\th[p,q]$ (denoted by $\bar{x}_j$) 
correspond to the poles of $y(x)$ with
negative residues. 
The values of $x$, where $y(x)=x$, correspond to the zeros of $ \f{d}{d q}\th[p+1/2,q+1/2]$.
 The values of $x$, where $y(x)=0$, correspond to the zeros of $ \f{d}{d q}\th[p+1/2,q]$.
And, finally, the  values of $x$, where $y(x)=1$, correspond to the zeros of $ \f{d}{d q}\th[p,q+1/2]$.
\end{Remark}

Consider, for example, the case of the negative cosmological constant (\ref{rc4}), when
$p\in\R$, $q=\f{1}{2}-i\tilde{q}$, $\tilde{q}>0$ (Hitchin's constants $k_1$ and $k_3$ are related to our
$p$ and $\tilde{q}$ as follows: $k_1=2p-1$, $k_3=2\tilde{q}$). Without loss of generality we can assume that
$\f{1}{2}< q \leq \f{3}{2}$. 

The zeros of $\th[p,q]$ can be easily found explicitly: these are 
\ben
\mu_n= \f{p+n-1/2}{\tilde{q}}\;,\hskip0.8cm n=1,2,\dots\;.
\een
It was proved in \cite{Hitc94} that zeros of
 $\f{d}{d q}\th[p,q]$ are situated between  zeroes of $\th[p,q]$:
$\bar{\mu}_j\in (\mu_j,\mu_{j+1})\;,\;\; j\in\Z$. If $1/2 \le p\leq 1$,  $\f{d}{d q}\th[p,q]$
does not have any other zeros; if  $1/2 \le p\le 3/2$ then there exists an additional zero $\bar{\mu}_0<\mu_1$.

It is easy to see that the singularities of the metric at $\mu_n$ are of the type of coordinate
singularity at the center of the unit ball.  In contrast, at the zeros of expressions
(\ref{sin3})-(\ref{sin5}) even the conformal structure of our metric is singular. 
It is the non-trivial fact
proved in (\cite{Hitc94}, p.92) that these zeros are absent inside of the intervals 
$(\bar{\mu}_n,\mu_{n+1}]$. Therefore, the metric (\ref{ggg}) on the interval $(\bar{\mu}_n,\mu_{n+1}]$
may be interpreted as the metric on the unit ball with the origin at $\mu=\mu_{n+1}$ and the boundary at
$\mu=\bar{\mu}_n$; the conformal structure of the metric (\ref{ggg}) obviously induces a  non-singular 
metric on the boundary sphere $S^3$.

{\bf Acknowledgements.} We thank Owen Dearricott who pointed out the wrong sign in the formula for $W_1$ in
the published version of this paper.

\end{document}